# Ferromagnetic metallic Sr-rich $Ln_{1/2}A_{1/2}CoO_3$ cobaltites with spontaneous spin rotation

Jessica Padilla-Pantoja[1], Arnau Romaguera[1], Xiaodong Zhang[1], Javier Herrero-Martín[2], Francois Fauth[2], Javier Blasco[3] and José Luis García-Muñoz[*1]

[1]*Institut de Ciència de Materials de Barcelona, ICMAB-CSIC, Campus UAB, 08193 Bellaterra, Spain*

[2]*CELLS-ALBA Synchrotron, 08290 Cerdanyola del Vallès, Barcelona, Spain*

[3]*Instituto de Nanociencia y Materiales de Aragón, Departamento de Física de la Materia Condensada, CSIC-Universidad de Zaragoza, C/ Pedro Cerbuna 12, 50009 Zaragoza, Spain*

**\* Corresponding autor:**
Prof. José Luis García-Muñoz
*E-mail:garcia.munoz@icmab.es*
*Institut de Ciència de Materials de Barcelona, ICMAB-CSIC, Campus universitario de la UAB, 08193 Bellaterra, Spain*

**Abstract**

The $Pr_{0.50}Sr_{0.50}CoO_3$ perovskite exhibits unique magnetostructural properties among the rest of ferromagnetic/metallic $Ln_{0.50}Sr_{0.50}CoO_3$ compounds. The sudden orthorhombic-tetragonal (*Imma*→*I4/mcm*) structural transition produces an unusual magnetic behavior versus temperature and external magnetic fields. In particular, the symmetry change is responsible for a spontaneous spin rotation in this metallic oxide. We have studied half-doped $Ln_{0.50}(Sr_{1-x}A_x)_{0.50}CoO_3$ cobaltites varying the ionic radius $r_A$ of A-site cations (divalent cations and lanthanides) in order to complete the T-$r_A$ phase diagram. The influence of the structural distortion and the A-cations size for the occurrence of a spontaneous spin reorientation in the metallic state has been investigated. As the reorientation of the magnetization is driven by the temperature induced collapse of the orthorhombic distortion, a careful investigation of the structural symmetry is presented varying the structural distortion of the Sr-rich half-doped cobaltites by means of both compositional and temperature changes. The region in the phase diagram of these perovskites where the phase of magnetic symmetry *Fm´m´m* replaces that of *Im'm'a* symmetry was determined in this family of ferromagnetic/metallic cobaltites. In that region the magnetization direction has rotated 45º degrees within the *a-b* plane with respect to the second.

# I. INTRODUCTION

Cobalt oxides attract much attention as they present a rich variety of interesting phenomena. Among other remarkable properties cobalt compounds are well suited for applications as mixed conductors, materials for solid oxides fuel cells, thermopower applications, room temperature (RT) ferromagnets, superconductivity, magnetoelectric materials, high temperature solar absorbers or materials for spintronics [1–6]. Moreover, the ability of cobalt oxides containing $Co^{3+}$ ions to adopt different spin states is a relevant feature due to the importance of the spin state of Co for electron mobility, the nature of magnetic coupling or the structural stability. The perovskite-type cobaltites $LnCoO_3$, $Ln_{1-x}Ca_xCoO_3$ and $LnBaCo_2O_{5+d}$ (Ln=Lanthanide) are good examples of systems exhibiting phase transitions triggered by spin state changes [7–11]. Examples of both ferromagnetic metals and insulators (of great potential for spintronic applications) are found among the rich ecosystem of cobalt oxides [6].

The double-exchange interaction (as in the widely studied manganites) exemplifies an effective mechanism to produce ferromagnetic metallic states in many cobaltites. The hybridization between $3d$ metal and $2p$ oxygen orbitals is more intense in cobalt oxides than in manganites favoring the delocalization of charges. Referent examples are the half-doped $Pr_{0.50}A_{0.50}CoO_3$ (A: Ca, Sr, Ba) compounds, where $Co^{3+}$ and $Co^{4+}$ species coexist in 1:1 ratio. In spite of sharing a metallic behavior, their properties and structure strongly differ.

(i) $PrBaCo_2O_6$ was assigned to the *P4/mmm* space group (SG) with ordered double perovskite structure in which Pr and Ba form alternating layers along the *c*-axis [12]. This "112" layered cobaltite presents a Curie temperature $T_C=210K$ in virtue of the double-exchange mechanism.

(ii) The half-doped cobaltite $Pr_{0.50}Ca_{0.50}CoO_3$ (PCCO) is metallic but undergoes a metal-insulator transition (MIT) upon cooling at $T_{MI} \sim 80$ K, becoming insulating [7]. Its *Pnma* crystal symmetry is preserved across the MIT [13]. This Ca compound does not exhibit spontaneous magnetic order but a exceptional electronic mechanism based on the stabilization of the $Co^{3+}$ LS state at $T_{MI}$. A spin-state change in a fraction of cobalt atoms is concurrent at $T_{MI}$ with a sudden partial $Pr^{3+}$ to $Pr^{4+}$ valence change. Consequently, at the MIT there is a transfer of electrons from Pr to Co sites [14–16]. $Pr_{0.50}Ca_{0.50}CoO_3$ attracts interest in the area of ultrafast optical switching devices due to the proven ability of generating metallic domains in the insulating low temperature phase by ultrafast photoexcitation [17–20].

(iii) The MIT of PCCO does not take place in $Pr_{0.50}Sr_{0.50}CoO_3$ (PSCO). The compound doped with Sr presents a (less distorted) *Imma* structure [21,22], where the partial $Pr^{3+}$ to $Pr^{4+}$ valence shift and Co spin-state transition of PCCO have been suppressed [23]. Instead, Mahendiran *et al.* [24] reported unexpected magnetic anomalies at $T_{SR} \sim 120$ K, well below the ferromagnetic (FM) transition at $T_C \sim 230$ K. A reorientation of the magnetization axis was suggested by Lorentz Transmission Electron Microscopy (LTEM) images by Uchida *et al.* in ref. [25] and by means of x-ray magnetic circular dichroism (XMCD) experiments [26,27], and confirmed by neutron diffraction data and symmetry analyses [28]. A rotation of the magnetization axis by 45º within the *a-b* plane takes place spontaneously in PSCO across the second magnetic transition observed in the magnetization curves. This Sr-rich perovskite retains its metallic character across the unexpected magnetostructural transition at $T_{SR}$.

The discovery of materials presenting spin-rotation (SR) transitions attracts substantial attention within the condensed matter physics community. In the fundamental plane it is of great interest to recognize the microscopic origins of these transitions for a better understanding of order and dynamic phenomena in classical and quantum magnets. In applied fields developing nanocomponents for spintronic devices, such as spin-torques, the possibility of using spins instead of charges as the variable state for computation or operation demands manipulating the spin orientation of nanomagnets to switch between logic or conductive states [29,30]. Spin rotations can be also used for the optical control of magnetically ordered materials. For instance, as a result of laser-induced heating and a subsequent spin reorientation phase transition, ultrafast excitations of quasi-ferromagnetic modes have been reported for canted antiferromagnets such as some $LnFeO_3$ oxides [31].

Given the increasing importance of ferromagnetic oxides presenting spin-rotation transitions, in this paper we investigate the stability of the SR transition reported for the compound PSCO with respect to the distortion of the perovskite structure in the family of half-doped Sr-rich $Ln_{0.50}A_{0.50}CoO_3$ cobaltites.

## II. EXPERIMENTAL DETAILS

Half doped cobaltites $Ln_{0.50}A_{0.50}CoO_3$ with compositions relatively close to the reference compound PSCO were prepared by solid-state reaction or sol-gel methods. First, $Ln_{0.50}Sr_{0.50}CoO_3$ perovskites with *Ln*=Pr, $Pr_{0.95}Tb_{0.05}$, Nd and Tb were prepared by solid-state reaction at high temperature under oxygen atmosphere. High purity $Co_3O_4$, $Pr_6O_{11}$ and corresponding $Ln_2O_3$ oxides were first dried at 1100 ºC. $SrCO_3$ was heated at 850 ºC, then up to 950 ºC to achieve decarbonation. The precursors were then mixed up, pressed into pellets and heated in oxygen atmosphere at 1000 ºC for 12 h, with a slow cooling. After gridding and pressing, the annealing procedure was repeated several times, the last one being performed at 1170 ºC (during 24h) under $O_2$ flow followed by a slow cooling (60 ºC/h). Secondly, several compositions of the type $Pr_{0.50}(Sr_{0.50-x}Ca_x)CoO_3$ (with $0.025 \leq x \leq 0.10$) were also prepared following a similar procedure but adding $CaCO_3$. In addition, a bigger alkaline-earth ion was used in the mixed half-doped cobaltites $Pr_{0.50}(Sr_{1-x}Ba_x)_{0.50}CoO_3$, where the $Sr^{2+}$ ion is progressively substituted by the bigger $Ba^{2+}$. Several compositions of this series (x= 0.025, 0.05 and 0.10) were prepared by a sol-gel method, using stoichiometric amounts of $Pr_6O_{11}$, Co-metal, $SrCoO_3$ and $BaCoO_3$. The mixture of reactives was dissolved in a solution of $HNO_3$ 1M, by adding citric acid and ethylene glycol. The pink solution was slowly evaporated leading to a brown resin which was first dried at 650 ºC for 6h. The obtained precursor was pressed into pellets and heated under $O_2$ atmosphere for several annealings in the range 1175-1200 ºC, with intermediate grindings and pressings. The final product was heated again at 800 ºC for 16h under the same atmosphere ($O_2$) and cooled slowly (60 ºC/h).

X-ray diffraction patterns were recorded at RT using a Siemens D-5000 diffractometer and Cu Kα radiation. Electrical transport (using the four-probe method) and magnetization measurements were performed using a Physical Properties Measuring System (PPMS) and a Superconducting Quantum Interferometer Device (SQUID) from Quantum Design. Our preliminary characterization confirmed well crystallized and single phased perovskites.
Neutron diffraction measurements were carried out at the high-flux reactor of the Institut Laue Langevin (Grenoble, France) using the high-intensity D20 ($\lambda$= 1.594 Å, 1.87 Å), D1B ($\lambda$=2.52 Å) and the high resolution D2B ($\lambda$ = 1.594 Å) diffractometers. Neutron powder diffraction (NPD) patterns of several compositions were recorded at variable temperatures using helium cryostats and a cryofurnace, within the temperature range 10K-450K. Moreover, synchrotron x-ray powder diffraction (SXRPD) measurements were performed on the BL04-MSPD beamline [32] of the ALBA Synchrotron Light Facility (Barcelona, Spain) using a wavelength, $\lambda$ = 0.41290(3) Å, which was determined by measuring a NIST standard silicon. The samples were loaded in borosilicate glass capillaries (diameters of 0.5 and 0.7 mm) and kept spinning during data acquisition. For selected compositions, patterns between 10 and 300 K were collected using a Dynaflow liquid He cryostat. Two detection systems, the MAD detection setup [13-channel multianalyzer detector] and the position-sensitive detector MYTHEN, were used. Structural and magnetic Rietveld refinements were made using the Fullprof program [33]. Neutron refinement of the oxygen occupation factors did not detect oxygen vacancies within an estimated error of ~3 %. Variable oxygen occupancies didn't improve the agreement factors in the neutron fits.

## III. RESULTS AND DISCUSSION

### III.1 – Half-doped Sr-rich cobaltites with spontaneous spin rotation

**III.1.1** $Pr_{0.50}Sr_{0.50}CoO_3$

The structural and magnetic properties of the PSCO sample used in this work have been exhaustively described in refs. [21,22,28]. In these previous reports the successive structural and magnetic phases were investigated and comprehensively described as a function of temperature (T). Fig. 1 plots the magnetization of the PSCO sample (after field cooling) under 10 Oe and 1 kOe. Concurrent with the spin reorientation, at 120K there is a loss/gain of magnetization depending on the value of the applied field: $H<H_{cr}$ or $H>H_{cr}$, respectively (critical field $H_{cr} \approx 300$ Oe). We call FM1 and FM2 to the distinct ferromagnetic phases below and above the transition at $T_{SR} \sim 120$ K. The orbital and spin components of the collinear Co moments rotate by 45º across the transition [28].

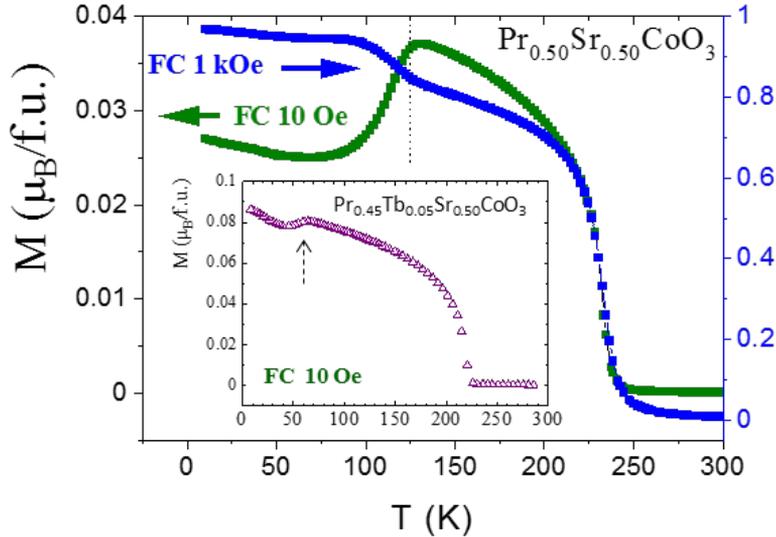

**Figure 1.** Field-cooled M(T) curves for PSCO showing negative and positive steps respectively under 10 Oe and 1 kOe. The SR transition is indicated as dotted line. Inset: FC M(T) for $Pr_{0.45}Tb_{0.05}Sr_{0.50}CoO_3$ (10 Oe), where we found $T_C \approx 224$ K and $T_{SR} \approx 60$ K.

Upon cooling from high temperatures, PSCO follows the succession of phase transformations $Pm-3m \rightarrow R-3c \rightarrow Imma \rightarrow I4/mcm$, all keeping the metallic state. The corresponding transition temperatures can be labelled as $T_{CR}$, $T_{RO}$ and $T_{OT}$, respectively. In the Glazer notation [34] the chain of changes between successive tilt systems can be easily visualized as $a^0a^0a^0 \rightarrow a-a-a- \rightarrow a-a-c^0 \rightarrow a^0a^0c-$. The two last transitions (at $T_{RO}=314$ K and $T_{OT}= T_{SR} =120$ K) are shown in Fig. 2(a), where we plot a projection of the thermal evolution of the neutron intensities (adapted from [21]). A comparison of the orthorhombic (O) and tetragonal (T) phases can be found in ref. [21], where selected interatomic distances and bond angles were also reported for the *Imma* (O) and *I4/mcm* (T) structures refined at 300 and 15 K.

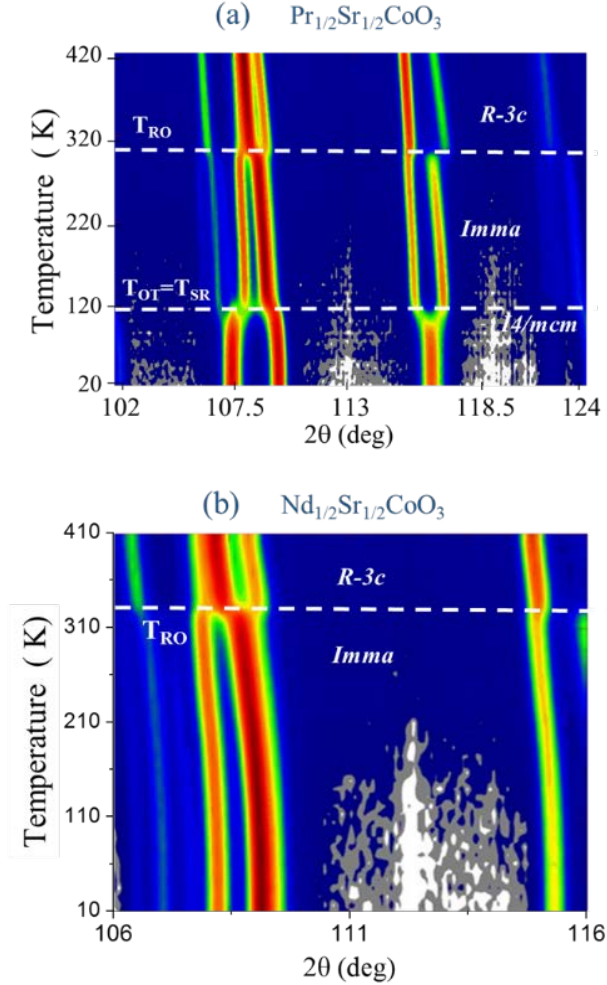

**Figure 2.** Neutron diffraction intensities for characteristic reflections of the successive crystal structures as a function of temperature (d20@ill, λ= 1.87 Å) for (a) $Pr_{0.50}Sr_{0.50}CoO_3$ [adapted from [21]] and (b) $Nd_{0.50}Sr_{0.50}CoO_3$. Notice the absence of the tetragonal phase in the Nd compound.

The characteristics of the FM1 and FM2 ferromagnetic phases were deduced by analyses of the NPD data and their respective magnetic space groups (MSG) were also determined by García-Muñoz *et al.* in ref. [28]. PSCO orders below the Curie point ($T_C$ = 230 K) with the ferromagnetic moments parallel to the *a* axis in the *Imma* cell ($F_x$, $\sqrt{2}a_0$ x $2a_0$ x $\sqrt{2}a_0$). We call this phase FM2 and its MSG is *Im'm'a* [Nr. 74.558, transformation to standard setting:(**a**, **b**, **c**; 0, 0, 0)]. At 140 K the refined FM moment was $m_x$=1.49(3) $\mu_B$/Co [28]. At 120 K the compound adopts the FM1 phase (SR) through a *Im'm'a* → *Fm'm'm* phase transition involving new structural and magnetic symmetries. As magnetic symmetry in the *I4/mcm* tetragonal phase ($\sqrt{2}a_0$ x $\sqrt{2}a_0$ x $2a_0$ cell) we found the *Fm'm'm* magnetic space group [Nr. 69.524, transformation to standard setting:(**-c, a-b, -a-b** ; 0, 1/2, 0)] with Co moments within the *a-b* plane and $m_x=m_y=1.32(3)$ μ$_B$/atom at 15 K ($F_{xy}$ or 'diagonal' ferromagnetism). This description gives perfect account of both the NPD and LTEM results [25,28].

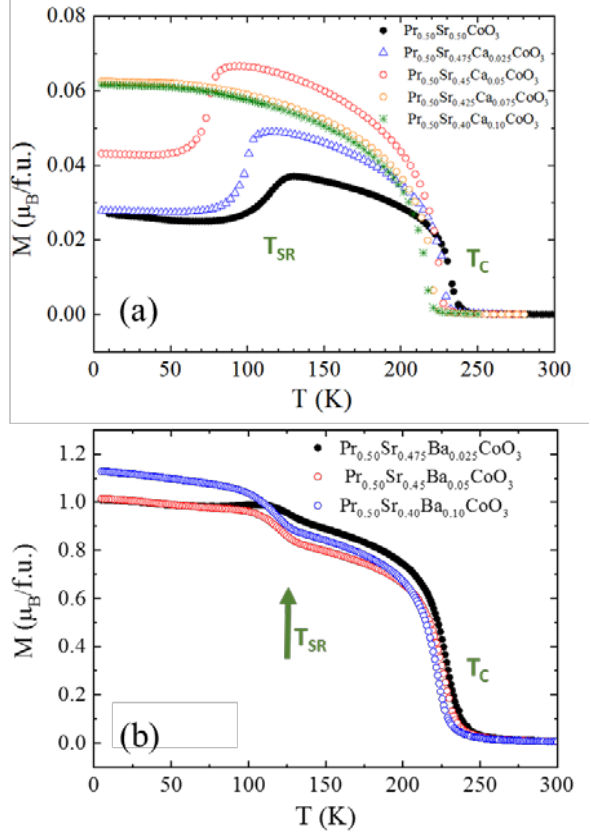

**Figure 3.** Magnetization (FC) for (a) Pr$_{0.50}$(Sr$_{1-x}$Ca$_x$)$_{0.50}$CoO$_3$ (x=0, 0.025, 0.05, 0.075 and 0.10; 10 Oe) and (b) Pr$_{0.50}$(Sr$_{1-x}$Ba$_x$)$_{0.50}$CoO$_3$ samples with Ba contents x=0.025, 0.05 and 0.10 (1 kOe). The step-like anomaly in M(T) discloses the occurrence of the 45º spin rotation (SR) of Co moments driven by the O-T symmetry change at T$_{SR}$. The SR anomaly is not present in Ca-7.5% and Ca-10% doped samples.

### III.1.2 Probing the spin-rotation in half-doped Pr$_{0.50}$(Sr$_{1-x}$A$_x$)$_{0.50}$CoO$_3$ (A=Ba, Ca).

The evolution of the properties observed in PSCO upon varying the average distortion of the perovskite structure was first investigated by means of magnetic measurements. Magnetization measurements were performed on two main sets of samples: Pr$_{0.50}$(Sr$_{1-x}$A$_x$)$_{0.50}$CoO$_3$ with (a) A=Ca to reduce the ionic radius at the A site, and (b) with A=Ba to increase the average radius $r_A$ and decrease the distortion of the perovskite structure. Magnetization and resistivity measurements bear out that all the compositions studied are FM and metallic. Figure 3 exposes the magnetization M(T) of the samples with Ca content x=0, 0.025, 0.05, 0.075 and 0.10 measured under field cooling (FC, 10 Oe ). The two magnetic transitions are visible as abrupt changes in the magnetization at $T_C$ (PM-FM2) and $T_{SR}$ (FM2-FM1) for the compositions x=0, 0.025 and 0.05. A single transition (at $T_C$) is observed in the two last Ca compositions, x[Ca]=0.075 and 0.10 indicating that the magnetostructural transition does not come about in these two perovskites ($Pr_{0.50}Sr_{0.425}Ca_{0.075}CoO_3$ and $Pr_{0.50}Sr_{0.40}Ca_{0.10}CoO_3$). The corresponding $T_C$ and $T_{SR}$ transition temperatures for the compositions doped with Ca can be seen in Table I. The presence of Ca produces a gradual decrease of the $T_{SR}$ values and the vanishing of this transition in the range x~6-7% Ca. A smoother decrease is observed for $T_C$ substituting Sr by Ca. Table I also shows the coercivity and remnant magnetization observed at 10 K, which was obtained from magnetic hysteresis loops collected with a maximum applied field of 70 kOe. Fig. S1 (Supplementary information [35]) shows the magnetic hysteresis cycles for the different compositions. The remnant magnetization is very similar, but there is a smooth decrease in the coercivity increasing the Ca substitution and suddenly a marked drop in the $H_c$ values for x[Ca]>0.07. This agrees with the results mentioned above for the Ca series and it is indicating that the ferromagnetic ground state for the compositions with x[Ca]≥0.075 is different (FM2) to the "tetragonal" ferromagnetic phase (FM1) of the samples with x[Ca]<0.075. For the latter, the evolution of the amplitude and sign of the magnetization jumps with the applied field is shown in Figure 4 (samples with x[Ca]=0 and 0.05).

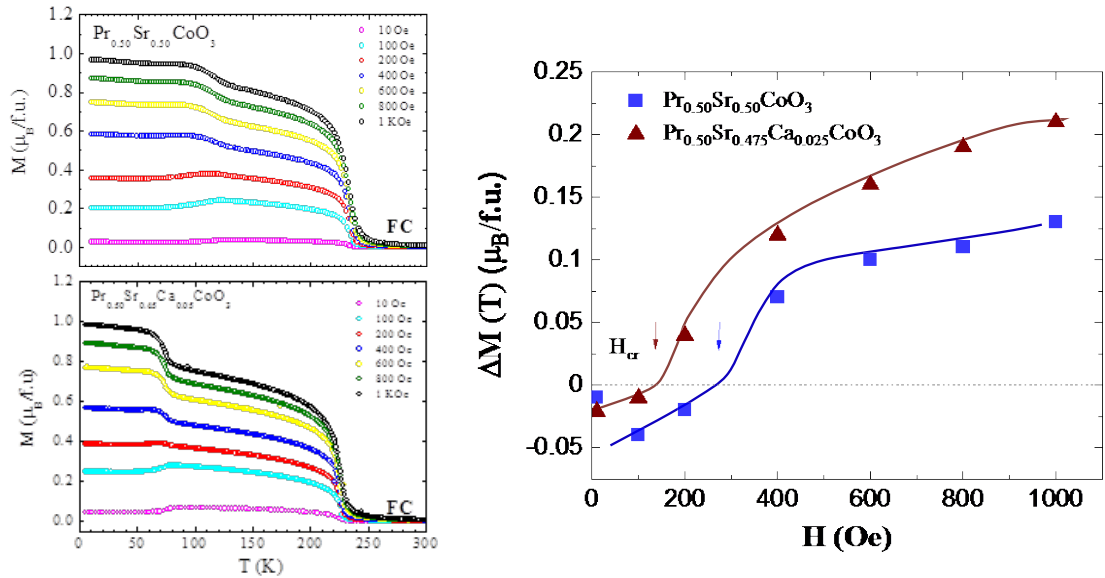

**Figure 4.** (left) Magnetization M(T) measured at different fields (H: 10, 100, 200, 400, 600, 800 and 1000 Oe, FC) in $Pr_{0.5}Sr_{0.5-x}Ca_xCoO_3$. Comparison for x=0 and x=0.05 samples. The sign of the magnetization jumps at $T_{SR}$ is field dependent. (right) Field evolution of the amplitude and sign of the magnetization jumps at the spin-rotation transition ($T_{SR}$ =$T_{OT}$). The field Hc for the crossover from negative to positive magnetization steps ΔM decreases from Hc~ 300 Oe (x=0) to 150 Oe (x=5% Ca).

In addition to the decrease of the FM2-FM1 (spin-reorientation) transition temperature ($T_{SR}$), Figure 4 shows an ostensible decrease in the critical field $H_{cr}$, from $H_{cr}$ ≈ 300 Oe (for x=0) to $H_{cr} \approx 150$ Oe (for x=0.05)). $H_{cr}$ defines the crossover from negative to positive ΔM values at $T_{SR}$. The loss of magnetization (negative step) under low fields (H< $H_{cr}$) is due to the presence of conjugated [110] and [1-10] magnetic domains below the *Imma* → *I4/mcm* transition. $H_{cr}$ is so related to the magnetic anisotropy within the *a-b* plane.

Independent confirmation of the suppression of the O-T structural transition in the samples with x[Ca]>0.07 was obtained by synchrotron diffraction on $Pr_{0.50}Sr_{0.40}Ca_{0.10}CoO_3$. This composition was found to adopt the orthorhombic *Imma* phase, without structural changes under cooling. We confirmed that this symmetry persists down to low temperatures, without being substituted by the tetragonal structure. Figure 5 shows the synchrotron diffraction pattern collected at 10 K for $Pr_{0.50}Sr_{0.40}Ca_{0.10}CoO_3$, satisfactorily refined using *Imma* symmetry. Its detailed *Imma* structure is reported in Table II as refined at 10 K. Therefore, its ground magnetic state corresponds to the ferromagnetic orthorhombic phase FM2. In contrast to $Pr_{0.50}Sr_{0.45}Ca_{0.05}CoO_3$, which exhibits a spin rotation at $T_{SR} \approx 79$ K and the tetragonal FM1 ground state.

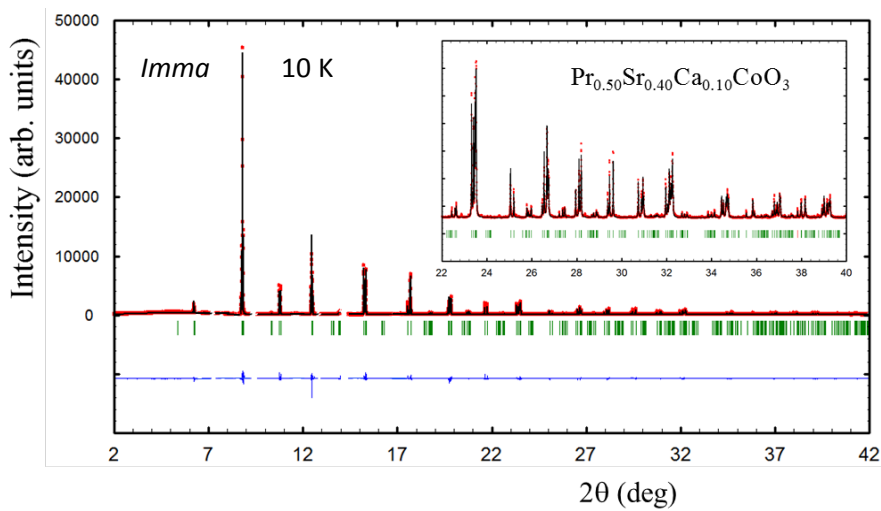

**Figure 5.** $Pr_{0.50}Sr_{0.40}Ca_{0.10}CoO_3$: Rietveld refinement of the synchrotron x-ray pattern at 10 K using the SG *Imma* (mspd@alba). Inset: detail of the high-angles region.

Next we will present the main results obtained when, instead of the smaller $Ca^{2+}$ ions, the larger $Ba^{2+}$ are partially replacing $Sr^{2+}$ sites. Single-phased well crystallized samples of the $Pr_{0.50}(Sr_{1-x}Ba_x)_{0.50}CoO_3$ series were obtained up to a 10% of substitution rate (samples with higher Ba rates presented secondary phases). Their x-ray diffraction patterns were well reproduced by the SG *Imma*. Figure 3(b) displays the T-evolution of the magnetization for three $Pr_{0.50}(Sr_{1-x}Ba_x)_{0.50}CoO_3$ compositions increasing the Ba content (x[Ba]=0.025, 0.05 and 0.10). It is observed that all these compositions exhibit the characteristic anomaly in M(T) produced by the occurrence of the O-T symmetry change and the reorientation of the magnetization ($T_{SR}$). $T_{SR}$ and $T_C$ transition temperatures deduced from the magnetization are given in the Table I for the compositions with 2.5%, 5% and 10% of Ba. From magnetic hysteresis loops recorded with a maximum applied field of 70 kOe (see Fig. S1 in the Supplementary information [35]), the coercivity and remnant magnetization observed at 10 K (FM1 magnetic phase) are also shown in Table I. As expected, both the remnant magnetization and the coercive field at 10 K are similar to the values found in the other compositions where we have confirmed a low-T $I4/mcm$ structure. It is shown that the coercivity in the ground state is systematically higher in the specimens keeping the $Imma$ symmetry.

**TABLE I.** Transition temperatures and magnetic properties obtained from magnetization measurements. The coercive fields ($H_c$) and remnant magnetizations ($M_r$) were obtained from magnetic hysteresis loops measured at 10 K.

| Sample | Low-T Structure | $T_C$ (K) | $T_{SR}$ (K) | $M_r$ ($\mu_B$/f.u) | $H_c$ (Oe) |
|---|---|---|---|---|---|
| $Pr_{0.50}Sr_{0.50}CoO_3$ | $I4/mcm$ | 236 | 124 | 0.605 | 466 |
| $Pr_{0.50}Sr_{0.475}Ba_{0.025}CoO_3$ | $I4/mcm$ | 235 | 129 | 0.596 | 395 |
| $Pr_{0.50}Sr_{0.45}Ba_{0.05}CoO_3$ | $I4/mcm$ | 231 | 128 | 0.594 | 390 |
| $Pr_{0.50}Sr_{0.40}Ba_{0.10}CoO_3$ | $I4/mcm$ | 226 | 126 | 0.597 | 160 |
| $Pr_{0.50}Sr_{0.475}Ca_{0.025}CoO_3$ | $I4/mcm$ | 229 | 100 | 0.614 | 381 |
| $Pr_{0.50}Sr_{0.45}Ca_{0.05}CoO_3$ | $I4/mcm$ | 227 | 79 | 0.677 | 377 |
| $Pr_{0.45}Tb_{0.05}Sr_{0.50}CoO_3$ | $I4/mcm$ | 224 | 60 | 0.907 | 483 |
| $Nd_{0.50}Sr_{0.50}CoO_3$ | $Imma$ | 225 | - | 0.651 | 1750 |
| $Pr_{0.50}Sr_{0.425}Ca_{0.075}CoO_3$ | $Imma$ | 221 | - | 0.652 | 842 |
| $Pr_{0.50}Sr_{0.40}Ca_{0.10}CoO_3$ | $Imma$ | 217 | - | 0.598 | 844 |
| $Tb_{0.50}Sr_{0.50}CoO_3$ | $Imma$ | 120 | - | 1.599 | 1832 |

**TABLE II**: Comparison of the crystal structures and reliability factors obtained at low temperature: from synchrotron x-ray (SXRPD, 10 K) and neutron patterns (NPD, 15 K). # mspd@alba ; *d20@ill (ref. 21)

|  | $Pr_{0.50}Sr_{0.40}Ba_{0.10}CoO_3$ | $Pr_{0.50}Sr_{0.50}CoO_3$ | $Pr_{0.50}Sr_{0.40}Ca_{0.10}CoO_3$ |
|---|---|---|---|
|  | **10 K (SXRPD)#** | **15 K (NPD)*** | **10 K (SXRPD)#** |
| Space group | *I 4/mcm* (140) | *I 4/mcm* (140) | *I mma* (74) |
| $a$ (Å) | 5.3732(1) | 5.3585(1) | 5.35768(3) |
| $b$ (Å) | 5.3732(1) | 5.3585(1) | 7.57005(4) |
| $c$ (Å) | 7.7053(1) | 7.7093(1) | 5.41025(3) |
| Vol (Å$^3$) | 222.463(5) | 221.367(9) | 219.428(2) |
| Pr/Sr/A | 4$b$ | 4$b$ | 4$e$ |
| $x$ | 0 | 0 | 0 |
| $y$ | 0.5 | 0.5 | 0.25 |
| $z$ | 0.25 | 0.25 | -0.0014(2) |
| $B$ (Å$^2$) | 0.35(2) | 0.790(6) | 0.342(7) |
| Co | 4$c$ | 4$c$ | 4$b$ |
| $x$ | 0 | 0 | 0 |
| $y$ | 0 | 0 | 0 |
| $z$ | 0 | 0 | 0.5 |
| $B$ (Å$^2$) | 0.11(2) | 0.459(2) | 0.176(11) |
| O1 | 4$a$ | 4$a$ | 4$e$ |
| $x$ | 0 | 0 | 0 |
| $y$ | 0 | 0 | 0.25 |
| $z$ | 0.25 | 0.25 | 0.4444(9) |
| $B$ (Å$^2$) | 1.66(15) | 1.190(1) | 0.91(12) |
| O2 | 8$h$ | 8$h$ | 8$g$ |
| $x$ | 0.2205(8) | 0.2849(3) | 0.25 |
| $y$ | 0.7205(8) | 0.7849(3) | 0.0242(7) |
| $z$ | 0 | 0 | 0.75 |
| $B$ (Å$^2$) | 1.26(11) | 1.028(6) | 1.65(8) |
| $\chi^2$ | 7.09 | 2.38 | 8.71 |
| $R_B$ (%) | 4.26 | 2.39 | 4.79 |
| $R_{mag}$ (%) | - | 7.2 | - |

The phase transitions in $Pr_{0.50}Sr_{0.40}Ba_{0.10}CoO_3$ were further investigated by synchrotron and neutron diffraction. It is the Ba doped (Sr-rich) cobaltite with the largest cell (and lowest distortion) that we synthetized. Diffraction data were collected within the range 2 K-300 K. Figure 6 displays a projection of the thermal evolution of the synchrotron x-ray diffracted intensities in selected angular intervals. One can see the O-T structural transition ($T_{OT}$) extending from around ~127 K down to around ~80 K, consistently with the anomalous step revealed by the magnetization ($T_{SR}$). In Fig. 7(b) we have plotted the thermal evolution of the coexisting orthorhombic and tetragonal structures during the O-T transformation. Table II reports the atomic coordinates refined for $Pr_{0.50}Sr_{0.40}Ba_{0.10}CoO_3$ at 10 K (*I4/mcm*). Figure 7(a) exposes the Rietveld refinement plot corresponding to the tetragonal low-temperature phase (at 10 K). A small residual amount of untransformed phase (~9(1) % wt.) was detected in the sample at low temperatures. The two structures are represented in Figure 7(c).

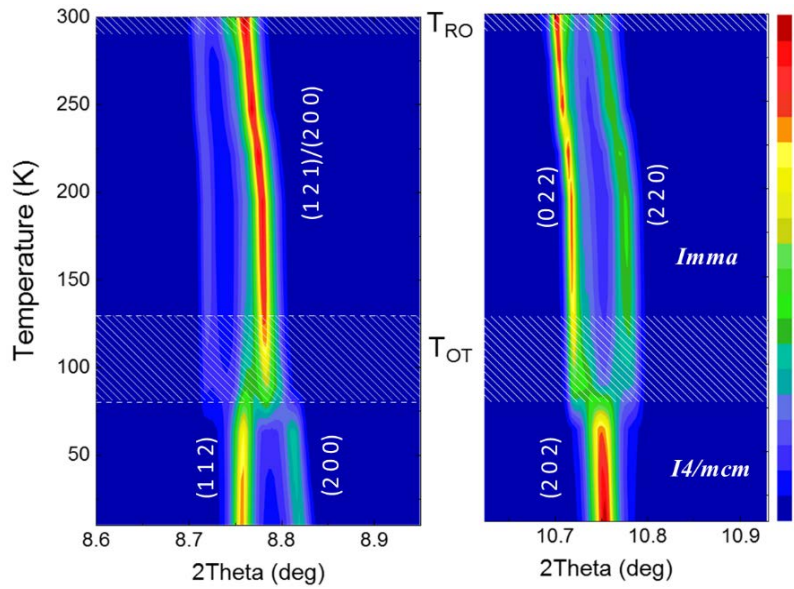

**Figure 6.** Pr0.50Sr0.40Ba0.10CoO3: evolution of the synchrotron x-ray diffraction intensities for characteristic reflections showing successive crystal structures as a function of temperature.

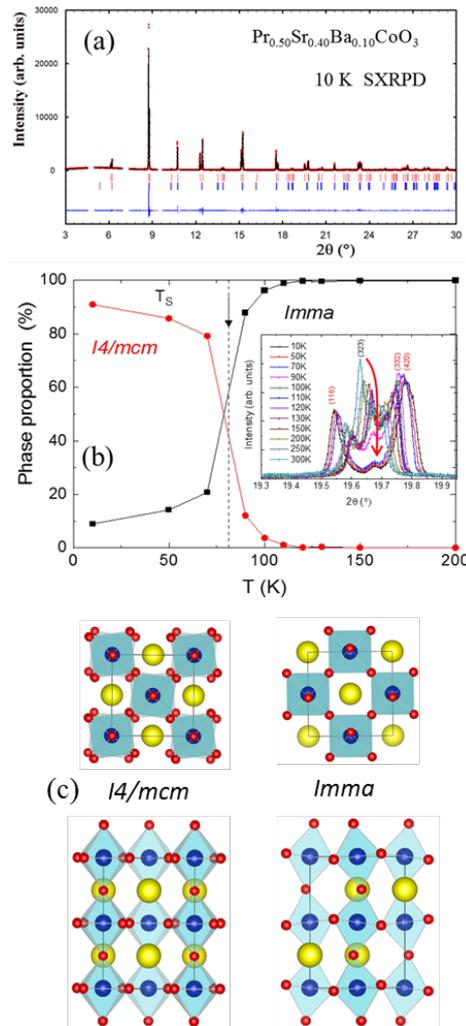

Figure 7. $Pr_{0.50}Sr_{0.40}Ba_{0.10}CoO_3$: (a) Rietveld refinement of the synchrotron x-ray pattern at 10 K (*I4/mcm*, mspd@alba). (b) T-evolution of the coexisting phase fractions across the low temperature transition. Inset: a small residual amount of untransformed orthorhombic phase is shown at 10 K [~9(1) % wt. of the O phase is included in (a)]. **(c)** Projections of the *Imma* and *I4/mcm* structures.

The *Imma* structure refined at 300 K is shown in Table III. This refinement evidenced that the O-R transition had already been initiated at 300 K. At this temperature, a minor fraction of the *R-3c* phase was included to fit the SXRPD pattern (≈19% wt.). The beginning of the O-R transition at $T_{OR}$~300 K upon heating is also suggested by the emerging changes distinguishable in Fig. 6 very near room temperature.

In addition, the evolution of the ferromagnetic order was studied using the high intensity, low resolution powder diffractometer D1B (λ=2.52 Å), well suited for magnetic order determination. A T−2θ projection of the thermal evolution of the main magnetic neutron-diffracted intensities is plotted in Figure 8 . Figure 9 plots the refinement of the neutron patterns collected at 285, 140 and 10 K for $Pr_{0.50}Sr_{0.40}Ba_{0.10}CoO_3$. At the two former temperatures the compound is *Imma*, and below $T_c$≈227 K it adopts the ferromagnetic order *Im'm'a* (magnetization parallel to *b* and 1.406(3) $\mu_B$/Co at 140 K, Figure 9(b)). Below the spin-rotation transition at $T_{SR}$~ 80 K , the *I4/mcm* structure was used for the neutron refinements using the *Fm'm'm* magnetic symmetry (FM1 phase) with $m_x=m_y$ ($F_{xy}$ with $m_{Co}$=1.78(3) $\mu_B$/Co moments along the diagonal of the *a-b* plane). The inset in Figure 9(c) presents a schematic view of the low-T ferromagnetic order. At 10 K the magnetic moment reaches a value of 1.78(3) $\mu_B$/Co, very similar to the moment found in PSCO ($m_{Co}$=1.87(4) $\mu_B$/Co [28]). The structural and magnetic details and reliability factors of the neutron refinements at 285, 140 and 10 K are listed in Table S1 (Supplementary information [35]).

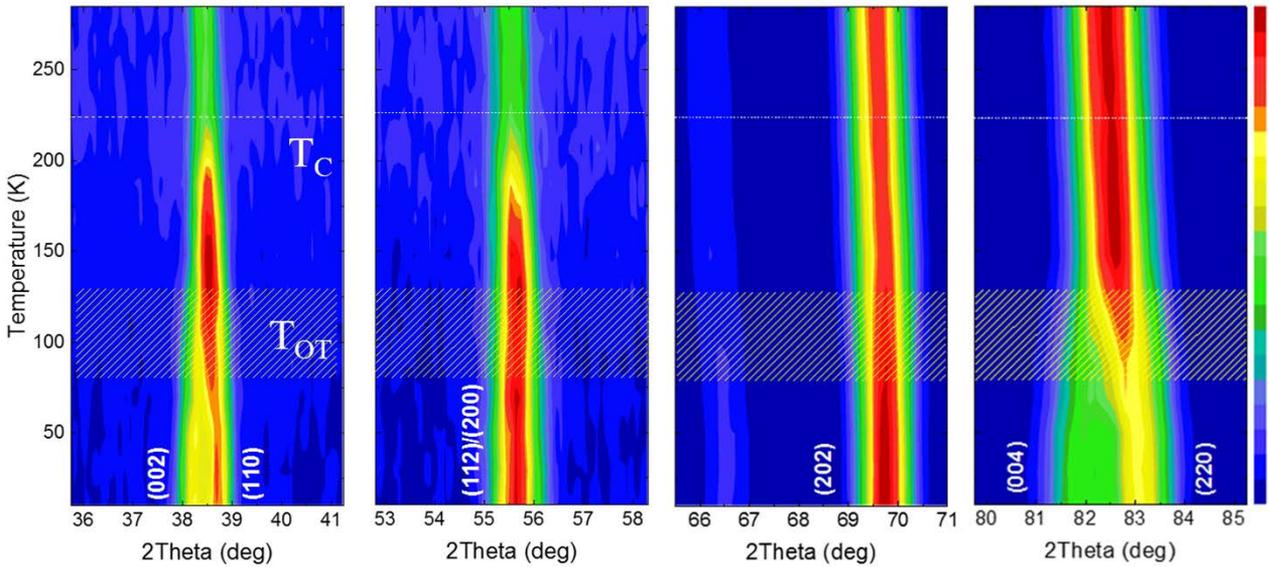

**Figure 8**. $Pr_{0.50}Sr_{0.40}Ba_{0.10}CoO_3$: T−2θ projection of the T-dependence for the neutron-diffracted intensities around selected reflections. Changes are apparent at Tc and $T_{SR}=T_{OT}$ (symmetry and spin-rotation transition, d1b@ill).

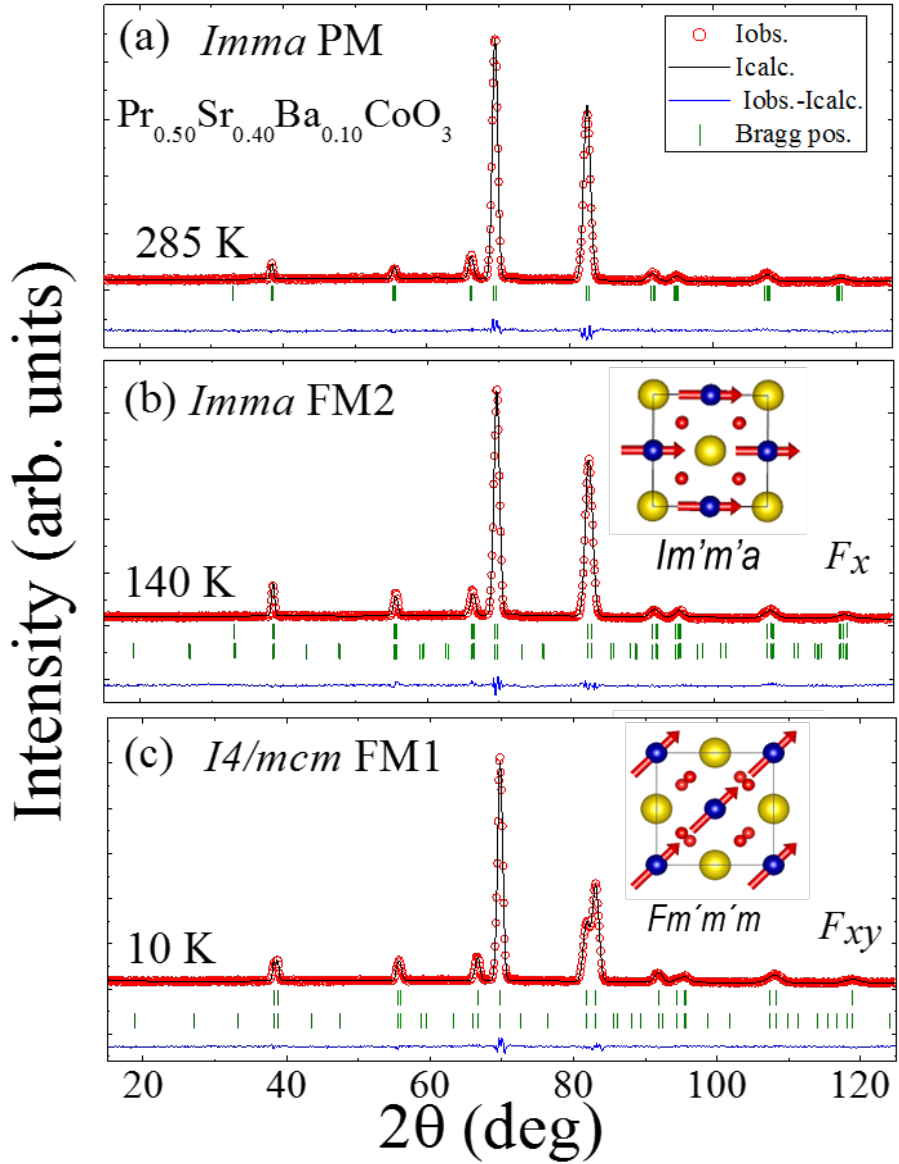

**Figure 9**. $Pr_{0.50}Sr_{0.40}Ba_{0.10}CoO_3$: Rietveld refinement of the neutron patterns (d1b@ill) obtained at (a) 285 K (*Imma* SG), (b) 140 K (*Im'm'a* MSG) and (c) 10 K (*Fm'm'm* MSG). A projection of the magnetic structures is also shown (yellow balls are Pr/Ba/Sr atoms, blue are Co atoms and O atoms are red).

The FM ordering temperature Tc hardly varies in the samples with Ba. Tc shows a very tiny decrease increasing Ba content ($Tc^{onset} \approx$ 238, 237, 231, 227 K for x=0, 2.5, 5 and 10%) very likely due to the increase of the A-cation size variance ($\sigma^2$), as occurs in similar FM perovskites with cation disorder [36,37]. Similarly, it is found that the variations observed in $T_{OT}$ under the Sr/Ba substitution are less pronounced than in the Sr/Ca substituted compounds.

**III.2 – Half-doped Sr-rich cobaltites without spontaneous spin rotation**

Magnetic measurements indicate that decreasing further the size of A-site cations only one ferromagnetic phase is preserved in these family of half doped cobaltites. Namely, the tetragonal phase and concurrent Co spin rotation do not occur, due to the stabilization of the *Imma* symmetry even at low temperatures (ground state). So, the magnetization and transport measurements reported by Yoshii *et al.* in refs. [38,39] revealed a single ferromagnetic transition (Tc) in $Ln_{0.50}Sr_{0.50}CoO_3$ perovskites for *Ln* = Nd, Sm, Eu or Gd. These four distorted half doped cobaltites are still metallic below room temperature in spite of their larger Co-O-Co distortions. A variety of space groups are found in the literature to describe these compounds due to the lack of more rigorous structural studies. For example, in refs. [39,40] the XRD patterns for $Ln_{0.50}Sr_{0.50}CoO_3$ samples were assumed to be *Pnma* (for Nd and Sm) and *Pm-3m* (for Eu). In contrast, the x-ray data of $Nd_{0.50}Sr_{0.50}CoO_3$ were refined using the *Imma* SG in ref. [41]. The *Pnma* symmetry was also attributed to the Nd compound in Refs. [38,42]. In ref. [43] the *Pnma* symmetry was discarded for $Eu_{0.50}Sr_{0.50}CoO_3$. The structure of $Tb_{0.5}Sr_{0.5}CoO_3$ was first refined by Srikiran *et al.* [44] as *Pnma* in all the temperature range below 300 K.

In view of this dispersion of descriptions, next we report a structural study of $Nd_{0.50}Sr_{0.50}CoO_3$ and the severely distorted $Tb_{0.5}Sr_{0.5}CoO_3$ cobaltite. Our analyses reveal that at RT both present *Imma* symmetry.

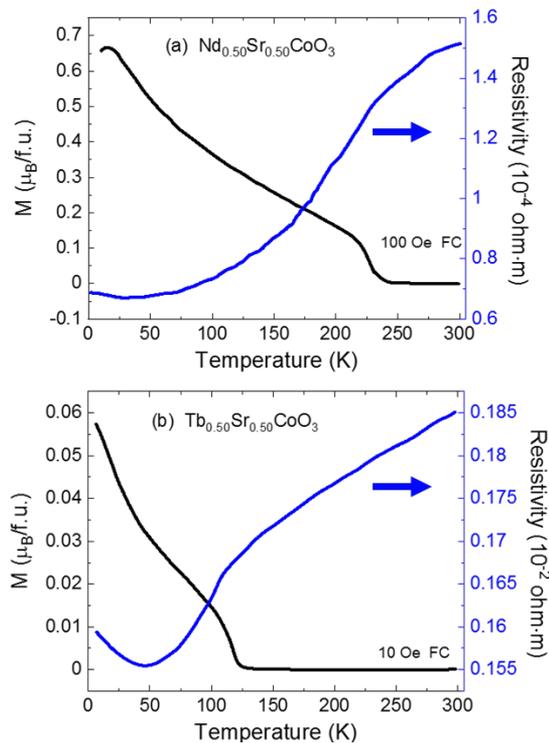

**Figure 10**. FC magnetization curves and resistivity of (a) $Nd_{0.50}Sr_{0.50}CoO_3$ and (b) $Tb_{0.50}Sr_{0.50}CoO_3$.

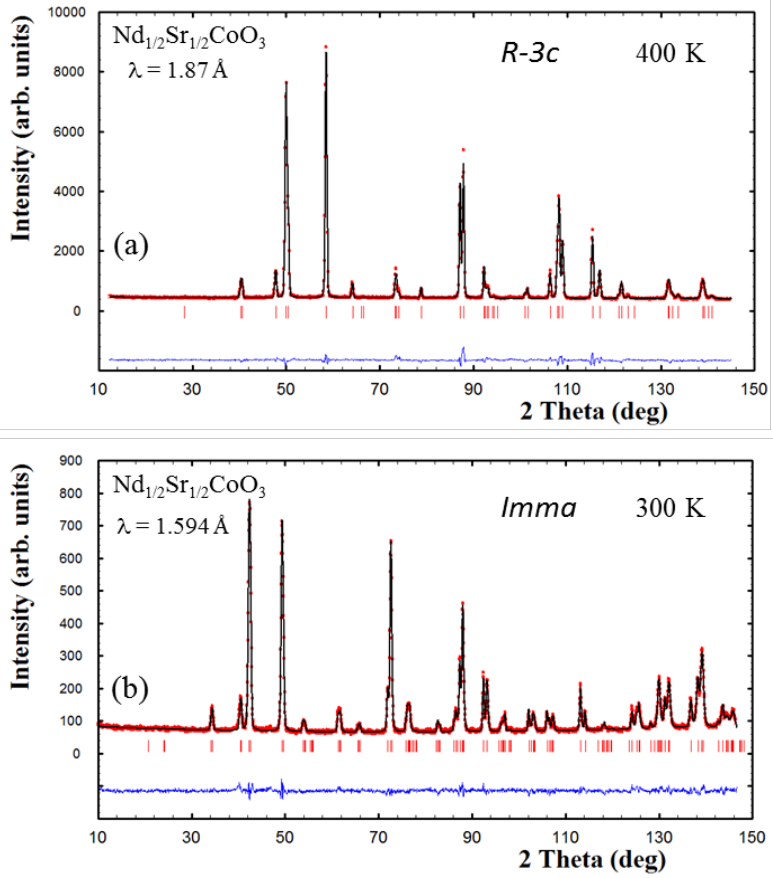

**Figure 11.** $Nd_{1/2}Sr_{1/2}CoO_3$: Rietveld refinements of the neutron patterns at (a) 400 K, *R-3c* (d20@ill) and (b) 300 K, *Imma* (d2b@ill).

*$Nd_{0.50}Sr_{0.50}CoO_3$*. Its magnetization and metallic resistivity below RT are exposed in Fig. 10. A drop in the magnetization curve was reported below ~100 K when the magnetization is measured under moderate *dc* fields (H ≥~1000 Oe) [38]. This behavior is not related with the SR and it comes from an AFM alignment between Nd(4*f*) and Co(3*d*) moments. Fig. 11(b) plots the refinement of a high-resolution NPD pattern at 300 K on this Nd cobaltite (d2b@ill, λ = 1.594 Å). One can see that, unlike most of earlier reports proposing a *Pnma* structure, our high-resolution neutron data is perfectly reproduced by the *Imma* SG. After a carefully inspection it is observed that the reflections permitted by the *Pnma* symmetry that are forbidden by the *Imma* SG are not present in the neutron pattern. This is the case for example of 'h + k + l = 2n+1' or 'h + l = 2n+1' type reflections, forbidden only in the *Imma* cell (see Fig. S2 of the Supplementary information [35]). Moreover, we have investigated the structural evolution of the $Nd_{0.50}Sr_{0.50}CoO_3$ cobaltite by neutron diffraction in the range 10 K-410 K (d20@ill, λ = 1.87 Å). In Fig. 2(b) the thermal evolution of the neutron diffraction intensities discloses a single structural transition within this range around 340 K. Under heating conditions this compound exhibits the *Imma* → *R-3c* transformation at $T_{OR}$=340 K. As expected $T_{OR}$ is thus slightly higher than for PSCO (314 K). In agreement with the evolution of the magnetization, the *Imma* structure persists down to the lowest temperature, confirming that Nd$_{0.50}$Sr$_{0.50}$CoO$_3$ does not adopt the tetragonal ground state of Pr$_{0.50}$Sr$_{0.50}$CoO$_3$. The FM moment refined at 100 K was 1.58(5) $\mu_B$/Co , with the same *Im'm'a* MSG and orientation than the FM2 phase of PSCO (m$_x$, referred to the $\sqrt{a_0}$ x $\sqrt{a_0}$ x 2$a_0$ cell). In Table III we detail the structural parameters for Nd$_{0.50}$Sr$_{0.50}$CoO$_3$ at 300 K (*Imma*) and at 400 K (R-3c).

**Table III:** Crystal structures of Nd$_{1/2}$Sr$_{1/2}$CoO$_3$ refined at 300 K [d2b@ILL] and 400 K [d20@ILL] (neutrons), and of Gd$_{0.50}$Sr$_{0.50}$CoO$_3$ at 300 K (synchrotron x-ray [mspd@alba]).

|  | Pr$_{0.50}$Sr$_{0.40}$Ba$_{0.10}$CoO$_3$ | Nd$_{0.50}$Sr$_{0.50}$CoO$_3$ | | Tb$_{0.50}$Sr$_{0.50}$CoO$_3$ |
|---|---|---|---|---|
|  | 300 K | 300 K | 400 K | 300 K |
| Space group | *I mma* (74) | *I mma* (74) | *R -3c* (167) | *I mma* (74) |
| a (Å) | 5.4036(1) | 5.3725(7) | 5.4245(1) | 5.36395(3) |
| b (Å) | 7.6325(2) | 7.6025(2) | 5.4245(1) | 7.58660(4) |
| c (Å) | 5.4334(1) | 5.4278(2) | 13.1628(4) | 5.36134(3) |
| Vol (Å$^3$) | 224.086(8) | 221.697(2) | 335.425(2) | 218.175(2) |
| Ln/Sr | 4*e* | 4*e* | 6*a* | 4*e* |
| x | 0 | 0 | 0 | 0 |
| y | 0.25 | 0.25 | 0 | 0.25 |
| z | 0.0000(6) | 0.0004(6) | 0.25 | -0.0002(7) |
| B (Å$^2$) | 0.87(2) | 0.791(5) | 1.219(6) | 1.21(1) |
| Co | 4*b* | 4*b* | 6*b* | 4*b* |
| x | 0 | 0 | 0 | 0 |
| y | 0 | 0 | 0 | 0 |
| z | 0.5 | 0.5 | 0 | 0.5 |
| B (Å$^2$) | 0.49(3) | 0.149(9) | 0.724(5) | 0.20(1) |
| O1 | 4*e* | 4*e* | 18*e* | 4*e* |
| x | 0 | 0 | 0.465(3) | 0 |
| y | 0.25 | 0.25 | 0 | 0.25 |
| z | 0.4569(36) | 0.4591(5) | 0.25 | 0.4548(24) |
| B (Å$^2$) | 4.7(4) | 1.562(7) | 1.567(5) | 5.2(1) |
| O2 | 8*g* | 8*g* |  | 8*g* |
| x | 0.25 | 0.25 | - | 0.25 |
| y | 0.0174(17) | 0.0243(3) | - | 0.0088(18) |
| z | 0.75 | 0.75 | - | 0.75 |
| B (Å$^2$) | 1.79(19) | 1.231(2) | - | 5.2(1) |
| R$_B$ (%) | 5.52 | 3.58 | 3.32 | 5.95 |
| R$_f$ (%) | 5.65 | 3.58 | 3.32 | 13.7 |
| $\chi^2$ | 9.22 | 1.67 | 2.76 | 5.54 |

*Tb$_{0.50}$Sr$_{0.50}$CoO$_3$*. As the most distorted Ln$_{0.50}$Sr$_{0.50}$CoO$_3$ compound of this study we have considered the Terbium cobaltite. Tb$^{3+}$ being smaller than Nd$^{3+}$, Sm$^{3+}$, Eu$^{3+}$ or Gd$^{3+}$. Fig. 10(b) shows the magnetization and resistivity curves of the Tb$_{0.50}$Sr$_{0.50}$CoO$_3$ compound below 300 K. In spite of the small ionic size of Tb$^{3+}$ (1.095 Å, IX-coordination [45]), the Tb$_{0.50}$Sr$_{0.50}$CoO$_3$ solid solution is metallic with Curie temperature $T_C$ =120 K (see also [44]). Its magnetization indicates absence of spin rotation below the ferromagnetic transition and therefore an orthorhombic ground-state. The detailed structure of Tb$_{0.5}$Sr$_{0.5}$CoO$_3$ was first reported by Srikiran et al. [44] - using NPD- as *Pnma* in all the temperature range below 300 K. In view of the present results we have re-examined the structure of $Tb_{0.5}Sr_{0.5}CoO_3$ by means of synchrotron x-rays. A quality SXRPD pattern recorded on the MSPD beamline of the ALBA Synchrotron is shown in Fig. 12. Intensities were collected in the angular interval 4º-55º using λ=0.424826 Å. The suitability of both *Pnma* and *Imma* SGs was tested and compared between them. From the best fits using these two descriptions we obtained the following agreement factors: (a) $R_B$ = 6.86%, $R_f$ = 16.0%, Rp= 10.8, Rwp: 13.9% and $\chi^2$ = 4.05%, for *Pnma*; and (b) $R_B$ = 5.95, Rp= 10.8, Rwp: 13.6, $R_f$ = 13.7% and $\chi^2$ = 5.54%, for the *Imma* SG. The *Imma* refined pattern is shown in Fig. 12. In Fig. S3 (Supplementary information [35]) we compare a selected interval of the two refinements with the focus in the low-intensity scale. Several extremely low-intensity and very wide reflections were detected in the SXRPD pattern that are characteristic of the *Pnma* symmetry ([023], [131], [115], etc). To notice is that they are markedly much less intense and much wider as compared to *Imma* reflections. As indicated in Fig. S3 these tiny reflections have around 200 times less intensity than normal ones. Consequently, the structure of $Tb_{0.5}Sr_{0.5}CoO_3$ should be regarded as *Imma*, ruling out a *Pnma* type distortion, even though very minority short-range *Pnma* distortions may exist (perhaps related to some kind of defects favored by the cation-size mismatch). The refined *Imma* structure of $Tb_{0.5}Sr_{0.5}CoO_3$ is reported in Table III. Disorder effects due to the high cation-size mismatch in this solid solution can be observed in form of high values of some refined thermal factors.

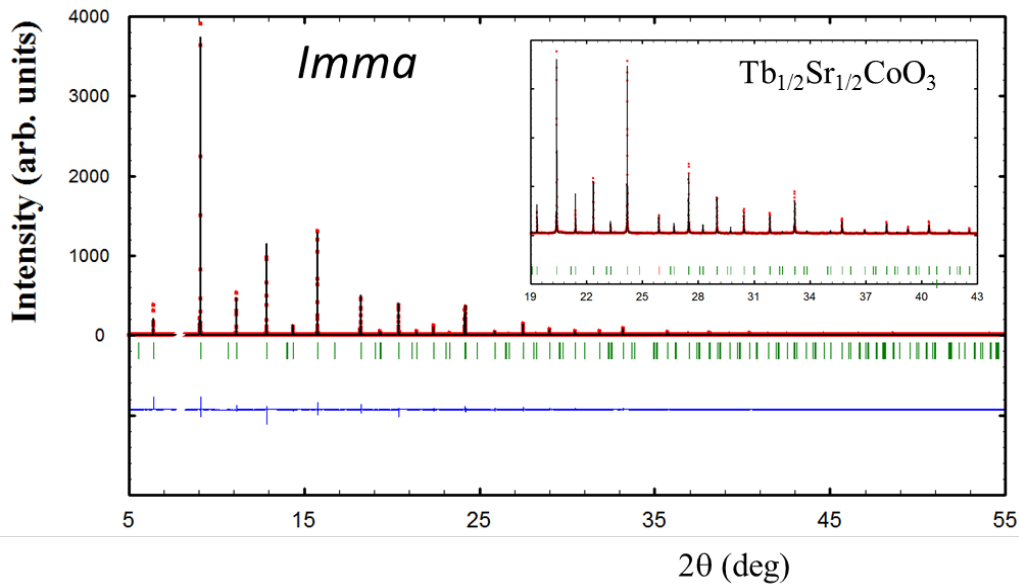

**Figure 12.** $Tb_{1/2}Sr_{1/2}CoO_3$: Rietveld refinement of the synchrotron x-ray pattern at 300K using the SG *Imma* (mspd@alba). Inset: detail of the 10º-30º interval.

*$Gd_{0.5}Sr_{0.5}CoO_3$*. Although $Gd_{0.5}Sr_{0.5}CoO_3$ was refined as *Pnma* using laboratory x-ray data in ref. [46], the size of Gd is in-between Nd and Tb ($\langle r_0 \rangle$: 1.163Å [$Nd^{3+}$] > 1.107Å [$Gd^{3+}$] > 1.095Å [$Tb^{3+}$]). As we have shown in the precedent paragraphs, the Nd and Tb half-doped cobaltites are both *Imma* below RT and therefore the Gd compound should also be considered of the same symmetry.

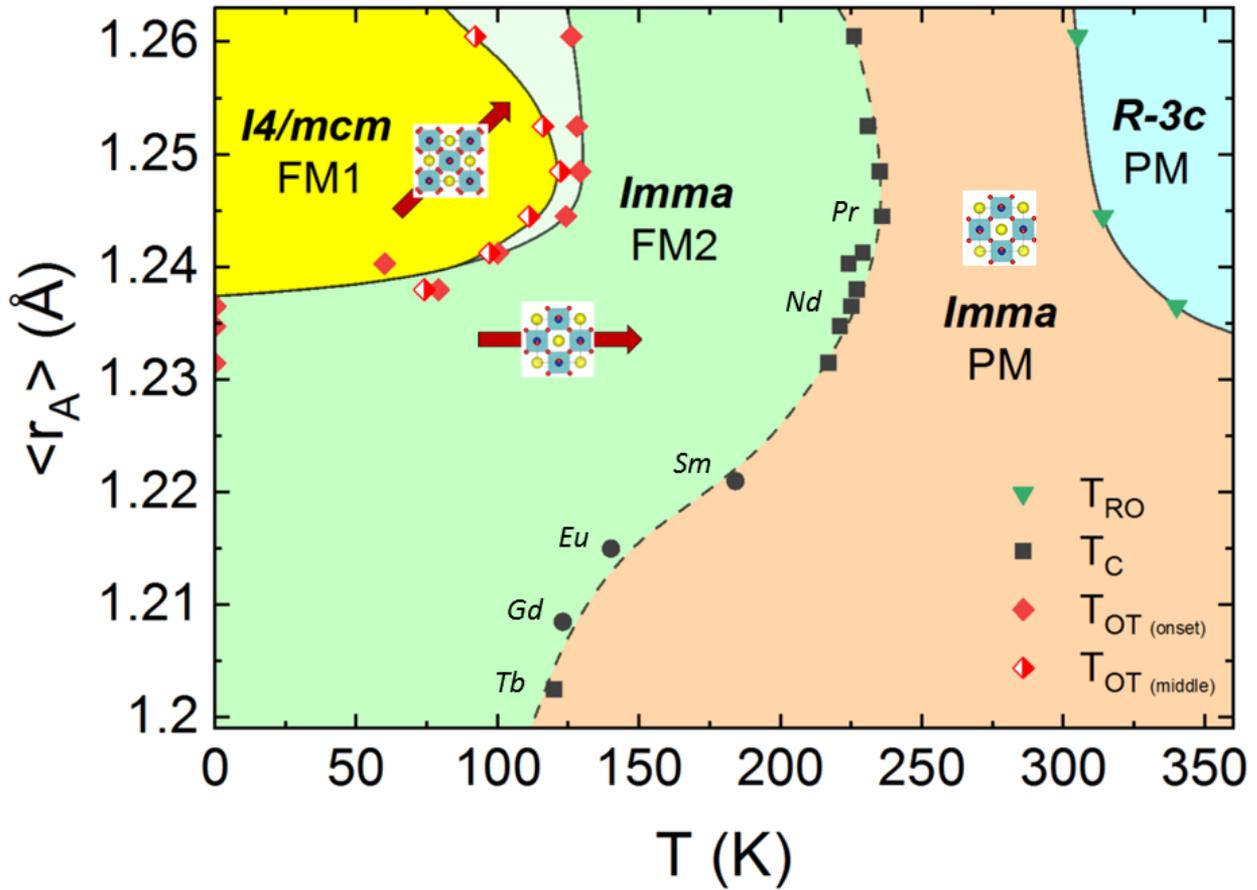

**Figure 13.** $\langle r_A \rangle$–T phase diagram of the half-doped Sr-rich $Ln_{0.50}(Sr, A)_{0.50}CoO_3$ cobaltites ($A$ = Ba or Ca). $T_{RO}$ are represented by green triangles (*R-3c*→*Imma*), the Curie temperatures $T_C$ are represented by black squares/circles (*PM*→*FM1*) and red rhombus represent $T_{OT}=T_{SR}$ (FM1/*Imma*→ FM2/*I4/mcm*). Black circles correspond to $T_C$ values for $Ln_{0.50}Sr_{0.50}CoO_3$ ($Ln$= Sm, Eu, Gd) extracted from Refs. [37,46,47]. The easy-axis in FM1 and FM2 phases are indicated.

### III.3 – Phase diagram

After the characterization and the results reported in the precedent sections, we have elaborated the $\langle r_A \rangle$–T phase diagram of the half-doped Sr-rich $Ln_{0.50}(Sr, A)_{0.50}CoO_3$ cobaltites shown in the Fig. 13. The different regions in the phase diagram correspond to: (i) paramagnetic R-3c ($a−a−a−$ distortion in Glazer notation); (ii) paramagnetic *Imma* ($a−a−c^0$ in the *Ibmm* setting [$\sqrt{2}a_0 \times \sqrt{2}a_0 \times 2a_0$], the setting used for the tetragonal cell); (iii) *Im'm'a* = Fx ferromagnetic (FM2) *Imma* phase; and (iv) *Fm´m´m* = Fxy ferromagnetic (FM1) *I4/mcm* phase ($a^0a^0c-$). The full numerical information gathered in Fig. 13 is also listed in Table S2 (Supplementary information [35]). $T_{RO}$ temperatures are represented by green triangles (*R-3c→Imma*), the Curie temperatures $T_C$ are represented by black squares/circles (*PM→FM1*) and $T_{SR}$ (=$T_{OT}$, onset) is represented by red rhombus (FM1/*Imma*→ FM2/*I4/mcm*). White&red rhombus indicate the middle point of the transition in the magnetization measurements. Black circles correspond to $T_C$ values for $Ln_{0.50}Sr_{0.50}CoO_3$ (*Ln*= Sm, Eu, Gd, extracted respectively from refs. [38,47,48]).

First, despite that distinct space groups were formerly used to describe the orthorhombic phases of $Ln_{0.50}Sr_{0.50}CoO_3$ compounds, Fig. 13 unveils that the orthorhombic structures are always *Imma*, at least down to *Ln*=Tb. Therefore, the rhombohedral-to-orthorhombic transition is of the type *R-3c* → *Imma*. So, none of the compounds studied showed *Pnma*.symmetry. Most interesting, there is a T-<$r_A$> confined region in the phase diagram of these Sr-rich half-doped cobaltites where the ground state exhibits a tetragonal *I4/mcm* structure and presents F$_{xy}$ magnetization from cobalt yielding overall *Fm´m´m* symmetry. This region has been determined and it is shown in Fig. 13. The borderline between this region and the *Im'm'a* phase identifies the line T$_{SR}$(<$r_A$>) of the phase diagram where the spontaneous spin rotation takes place. The reorientation of the easy-axis of cobalt follows to the suppression of the orthorhombic tilting of the octahedra in antiphase along *a* and *b* pseudocubic axes ($a-a-c^0$ → $a^0a^0c-$). The small tilting around the vertical axis that occurs in antiphase for successive octahedra along *c* is an additional distortion that does not play a main active role for the SR. The stability region of the *Fm´m´m* tetragonal magnetic phase vanishes for structural distortions very close to $Nd_{0.50}Sr_{0.50}CoO_3$. We found that the SR is suppressed in this cobaltite, which presents already an orthorhombic metallic ground state.

## 4. Conclusions.

The $Pr_{0.50}Sr_{0.50}CoO_3$ perovskite exhibits unique magnetostructural properties among the rest of ferromagnetic/metallic $Ln_{0.50}Sr_{0.50}CoO_3$ compounds. They are triggered by the transformation of an orthorhombic *Imma* structure into a tetragonal *I4/mcm* symmetry.
Summarizing, the stability of the ground ferromagnetic/metallic tetragonal structure (MSG *Fm´m´m*) has been investigated in bigger and smaller cells, partially substituting $Sr^{2+}$ by bigger ($Ba^{2+}$) and smaller ($Ca^{2+}$) alkaline-earth ions. In addition $Pr^{3+}$ ions were substituted by other lanthanides. Their structural and magnetic properties were characterized by synchrotron, neutron powder diffraction and magnetometry. As a result, we have built a T-$<r_A>$ phase diagram for half-doped $Ln_{0.50}(Sr_{1-x}A_x)_{0.50}CoO_3$ cobaltites below 360 K as a function of temperature and the distortion of the perovskite structures. We have identified the stability region where a tetragonal *Fm´m´m* metallic phase, with the magnetization direction rotated by 45º degrees within the *a-b* plane, replaces the orthorhombic *Im'm'a* phase with Fx ferromagnetic moments. The limits of this region have been determined and its boundaries draw the line where a spontaneous spin rotation of Co moments takes place.

**Acknowledgements.**

We acknowledge financial support from the Spanish Ministerio de Ciencia, Innovación y Universidades (MINCIU), through Projects No. RTI2018-098537-B-C21 and RTI2018-098537-B-C22, cofunded by ERDF from EU, and "Severo Ochoa" Programme for Centres of Excellence in R&D (FUNFUTURE (CEX2019-000917-S)). X.Z. was financially supported by China Scholarship Council (CSC) with No. 201706080017. A.R. and X.Z's work was done as a part of the Ph.D program in Materials Science at Universitat Autònoma de Barcelona. We also acknowledge ALBA (2014071047, 2015091512), ILL and D1B-CRG (MINCIU) for provision of beam time (CRG-D1B-14-293, CRG-2149-142, 5-31-1963).